\documentclass[
    twocolumn,
    superscriptaddress,
    amsmath,
    amssymb,
    aps,
    prb,
]{revtex4-2}

\usepackage{graphicx}
\usepackage{dcolumn}
\usepackage{bm}
\usepackage{mathtools}
\usepackage{physics2}
    \usephysicsmodule{ab}
    \usephysicsmodule{ab.braket}
    \usephysicsmodule{diagmat}
    \newcommand{\mqty}[1]{\begin{matrix}#1\end{matrix}}
    \usepackage{fixdif, derivative}
\usepackage[version=4]{mhchem}
\usepackage{here}
\usepackage{xcolor}
\definecolor{hrefcolor}{HTML}{039393}
\usepackage[colorlinks=true, allcolors=hrefcolor,]{hyperref}

\begin{document}
\preprint{APS/123-QED}

\title{Quantum-geometric origin of superfluid weight in quasicrystals with critical states}
\author{Kazuma Saito}
\affiliation{Department of Applied Physics, Tokyo University of Science, Katsushika, Tokyo 125-8585, Japan}

\author{Ryo Okugawa}
\affiliation{Department of Applied Physics, Tokyo University of Science, Katsushika, Tokyo 125-8585, Japan}

\author{Yusuke Kato}
\affiliation{Department of Basic Science, The University of Tokyo, Meguro, Tokyo 153-8902, Japan}
\affiliation{Department of Physics, The University of Tokyo, Bunkyo-ku, Tokyo, 113-0033, Japan}

\author{Takami Tohyama}
\affiliation{Department of Applied Physics, Tokyo University of Science, Katsushika, Tokyo 125-8585, Japan}

\date{\today}

\begin{abstract}
A distinctive feature of many quasiperiodic systems is the presence of critical states that are neither extended nor exponentially localized.
We investigate the geometric effect on the superfluid weight in quasiperiodic systems with critical states at zero temperature.
We employ both real-space and momentum-space approaches to superfluid weight in quasicrystals, which allows us to separate the conventional and quantum geometric contributions.
We find that the superfluid weight is dominated by the geometric contribution in quasiperiodic systems with critical states.
This finding reveals a fundamental interplay between superconductivity and critical states in quasicrystals.
\end{abstract}

\maketitle
The geometry of the quantum states is characterized by the quantum geometric tensor~\cite{Provost1980}.
In band geometry, its real part quantifies the distance between Bloch states, which is called the quantum metric.
Although Berry curvature effects on electronic states have been extensively studied~\cite{Xiao2010}, the quantum metric has only recently attracted attention in condensed matter physics~\cite{Resta2011, Rossi2021, Torma2022, Nagaosa2024}.
In superconductivity, the quantum metric affects superfluid weight
\cite{Peotta2015,Aleksi2016, Tovmasyan2016,Liang2017,Liang2017-2,Torma2018,Hu2019,Julku2020,Fang2020,Herzog2022,Kitamura2022FeSe,Kitamura2022FFLO,Alexander2022,Huhtinen2022,Mao2023,Kitamura2023,Porlles2023,Kitamura2024triplet,Chen2024,Mao2024,Kitamura2024pair,Hu2025,Porlles2025},
which characterizes Meissner effect.
In conventional theory, the superfluid weight is proportional to the inverse of the effective mass of a band, which corresponds to the intraband contribution to the supercurrent. 
In contrast, the geometric effect appears through the interband contribution of Bloch states to the supercurrent in multiband superconductors.
In this respect, the geometric interband contribution to the superfluid weight is different from the conventional one.
More recently, the superfluid weight in magic-angle twisted bilayer graphene has been experimentally observed~\cite{Tanaka2025}, and its temperature dependence, 
which deviates from the conventional theory,
indicates the presence of geometric effects~\cite{hirobe2025}.

The geometric contribution to superfluid weight can be dominant in topological flat-band systems because the quantum metric is enhanced due to the band topology.
This can also be understood in view of the localization property of the Wannier function~\cite{Peotta2015, Tovmasyan2016}.
When a flat band is topological with a nonzero Chern number, the Wannier functions can exhibit power-law decay rather than exponential decay~\cite{Marzari2012}.
The Wannier functions can easily overlap each other.
In this case, the interaction allows particles to move via the overlap,
which leads to a supercurrent. 
As a result, the superfluid weight can be finite in flat-band systems even if the conventional contribution is suppressed.
    
Quasiperiodic systems, such as quasicrystals~\cite{Shechtman1984,Levine1986}, lack periodicity but possess a long-range order.
While a momentum-space picture cannot be employed,
possible superconductivity has been theoretically investigated ~\cite{Sakai2017,Araujo2019, Sakai2019, Takemori2020, Cao2020, Liu2024, Saito2024}.
Also, superconductivity has been experimentally discovered in some quasicrystals~\cite{Kamiya2018,Tokumoto2024,Terashima2024}.
In quasicrystals,
some electronic states are neither extended nor exponentially localized,
which are called critical states~\cite{Ostlund1983,Tsunetsugu1986, Sutherland1986, Sutherland1987,Tokihiro1988, Tsunetsugu1991, Tsunetugu1991-b}. 
Despite the presence of critical states,
their role in superconductivity has remained elusive in quasicrystals.

Since the geometric effect on the superfluid weight is linked to the localization property of Wannier functions in periodic crystals,
the existence of critical states may give rise to a superfluid weight that cannot be understood from conventional theory.
It is therefore natural to ask whether a geometric effect on superfluid weights arises in quasicrystals.
In quasicrystals with isolated confined states, which lead to a sharp peak of the density of states (DOS), 
the geometric superfluid weight has been studied
in analogy with flat-band systems~\cite{Sun2025}.
However, it remains unclear whether the geometric contribution to the superfluid weight is enhanced in more general situations, where critical states are dominant.

In this Letter, we investigate the superfluid weight at zero temperature in quasicrystalline systems.
We employ two complementary approaches for superfluid weight in quasicrystals: a real-space formulation under open boundary conditions and a momentum-space one under periodic boundary conditions.
These approaches enable us to compare the conventional and quantum geometric contributions even in aperiodic systems.
We find that the critical states inherent to quasiperiodic systems significantly contribute to the superfluid weight via the quantum geometry.
Our results demonstrate that quantum geometry can provide a primary contribution to the superfluid weight in quasicrystals.

\textit{Superfluid weight.}
We first formulate the superfluid weight on both real-space and momentum-space basis. 
We start with the Hubbard model given by
\begin{align}
    \hat{H} = \sum_{\sigma} \sum_{j,l} (t_{jl} - \mu \delta_{jl}) \hat{c}_{j\sigma}^{\dagger} \hat{c}_{l\sigma} + U \sum_{j} \hat{c}_{j\uparrow}^{\dagger} \hat{c}_{j\downarrow}^{\dagger} \hat{c}_{j\downarrow} \hat{c}_{j\uparrow},
    \label{eq:Hubbard}
\end{align}
with $\hat{c}_{j\sigma}^{\dagger}$ ($\hat{c}_{j\sigma}$), the creation (annihilation) operator of an electron with spin $\sigma$ at site $j$.
Here, $t_{jl}$ is the hopping amplitude from site $l$ to $j$, $\mu$ is the chemical potential, and $U < 0$ is the onsite attractive interaction.
We consider $s$-wave superconductivity with time-reversal symmetry.

In the mean-field theory, the Bogoliubov-de Gennes (BdG) Hamiltonian can be written as
\begin{align}
    \hat{H} &= \sum_{j,l} \ab(\mqty{\hat{c}_{j\uparrow}^{\dagger} & \hat{c}_{l
    \downarrow}}) \mathcal{H}_{\mathrm{BdG}} \ab(\mqty{\hat{c}_{l \uparrow} \\ \hat{c}_{j\downarrow}^{\dagger}}),
    \\
    \mathcal{H}_{\mathrm{BdG}} &= \ab(\mqty{\mathcal{H} - \mathcal{V} & \bm{\Delta} \\ \bm{\Delta}^{\dagger} & -\mathcal{H}^{\ast} + \mathcal{V}}),
    \label{eq:H_BdG}
\end{align}
where $\bm{\Delta}$ is an order-parameter matrix where $\bm{\Delta}_{jl} = \Delta_{j} \delta_{jl}$ and $\Delta_{j} = U \braket<\hat{c}_{j\downarrow} \hat{c}_{j\uparrow}>$,
and $\mathcal{H}$ represents a tight-binding Hamiltonian for electrons in the normal phase.
$\mathcal{V}$ is the Hartree potential, where $\mathcal{V}_{jl} = V_{j} \delta_{jl}$, and $V_j = U\braket*<c_{j\uparrow}^\dagger c_{j\uparrow}> = U\braket*<c_{j\downarrow}^\dagger c_{j\downarrow}>$.

The superfluid weight $D_{\mu\nu} (\mu, \nu =x,y,z)$ can be calculated as the response of the supercurrent to an external vector potential.
Although supercurrents have been studied even in quasicrystals~\cite{Cao2020, Liu2022, Liu2023, Fukushima2023}, the geometric effect on the superfluid weight has not been clarified.
First, we derive the superfluid weight from the Kubo formula, based on the real-space picture under open boundary condition.
The detailed derivation is shown in Supplemental Material~\cite{SupplementalMaterial}.
Let $\ket|\psi_{a}>$ be an eigenvector of the BdG Hamiltonian in Eq.~\eqref{eq:H_BdG} with the energy eigenvalue $E_a$.
The superfluid weight is obtained from
\begin{align}
    D_{\mu\nu} &= \frac{1}{N} \sum_{a, b} \frac{f(E_a) - f(E_b)}{E_a - E_b}
    \nonumber \\
    &\times (\braket<\psi_a|{j}_\mu^p|\psi_b> \braket<\psi_b|{j}_\nu^p|\psi_a>
    \nonumber \\
    &- \braket<\psi_a|{j}_\mu^p \gamma_z|\psi_b> \braket<\psi_b|{j}_\nu^p \gamma_z|\psi_a>),
    \label{eq:D_munu in BdGbasis}
\end{align}
where ${j}_\mu^p$ is the matrix representation of the paramagnetic current operator $\hat{j}_{\mu}^p = -\delta \hat{H}/ \delta A_\mu$ with the $\mu$ component of the vector potential $A_\mu$.
Note that if $E_a=E_b$, we treat $\frac{f(E_a)-f(E_b)}{E_a-E_b}$ as $\partial f(E)/\partial E|_{E = E_a}$.
In Eq.~\eqref{eq:D_munu in BdGbasis}, $N$ is the system size and $f(E_a)$ is the Fermi distribution function at energy $E_a$, and $\gamma_z = \mathrm{diag}(1,-1) \otimes 1_{N}$ with the $N\times N$ identity matrix $1_N$.

We also consider superfluid weight for periodic systems to elucidate the geometric effect.
For periodic systems with the system size $N$, the superfluid weight is given by
\cite{Liang2017, Kitamura2022FeSe, Huhtinen2022, Kitamura2022FFLO}
\begin{align}
    &D_{\mu\nu}^{\mathrm{PBC}} = \frac{2}{N} \sum_{\bm{k},n,m,l,s} C_{nmls}^{++--}(\bm{k}) ([j_{\mu}^{p\uparrow}(\bm{k})]_{nm} \ab[j_{\nu}^{p\downarrow}(-\bm{k})]_{ls}
    \nonumber \\
    &\quad \quad \quad + (\mu \leftrightarrow \nu)),
    \label{eq:Dmunu from k space} \\
    &C_{nmls}^{++--}(\bm{k}) = \sum_{a, b} \frac{f(E_{a \bm{k}}) - f(E_{b \bm{k}})}{E_{a \bm{k}} - E_{b \bm{k}}} \phi_{n \bm{k}}^{a+\ast} \phi_{m \bm{k}}^{b+} \phi_{l \bm{k}}^{b-\ast} \phi_{s \bm{k}}^{a+}.
    \label{eq:Cnmls++--}
\end{align}
Here, $\phi_{n \bm{k}}^{a\sigma}$ are the expansion coefficients of the eigenstate $\ket|\psi_a>$ in terms of the product states of the normal-phase eigenstate $\ket|n\bm{k}>$ and the particle-hole state $\ket|\sigma>$.
Among these, the intraband contributions ($n = m$ and $l = s$) are interpreted as the conventional contribution $D_{\mu\nu,\mathrm{conv}}$, while the other interband contributions are interpreted as the quantum geometric contribution $D_{\mu\nu,\mathrm{geom}}$~\cite{Kitamura2022FFLO,hu2024}.
The geometric contribution is given as follows:
\begin{align}
    D_{\mu \nu, \mathrm{geom}}&=D_{\mu\nu}^{\mathrm{PBC}}-D_{\mu \nu, \mathrm{conv}}, \\
    D_{\mu \nu, \mathrm{conv}} 
    &=\frac{2}{N} \sum_{\bm{k},n m} C_{n n m m}^{++--}(\bm{k})
    ( [j_{\mu}^{p\uparrow}(\bm{k})]_{n n}[j_{\nu}^{p\downarrow}(-\bm{k})]_{m m} \notag \\ 
    &+ (\mu \leftrightarrow \nu)) ,
\end{align}
The details of the expression are provided in the Supplemental Material~\cite{SupplementalMaterial}.
Unlike $D_{\mu \nu, \mathrm{conv}}$,
$D_{\mu \nu, \mathrm{geom}}$ exhibits a strong dependence on the interaction~\cite{Peotta2015, Tovmasyan2016, Aleksi2016, Julku2020, Huhtinen2022}.

\begin{figure}
    \centering
    \includegraphics[width=\linewidth]{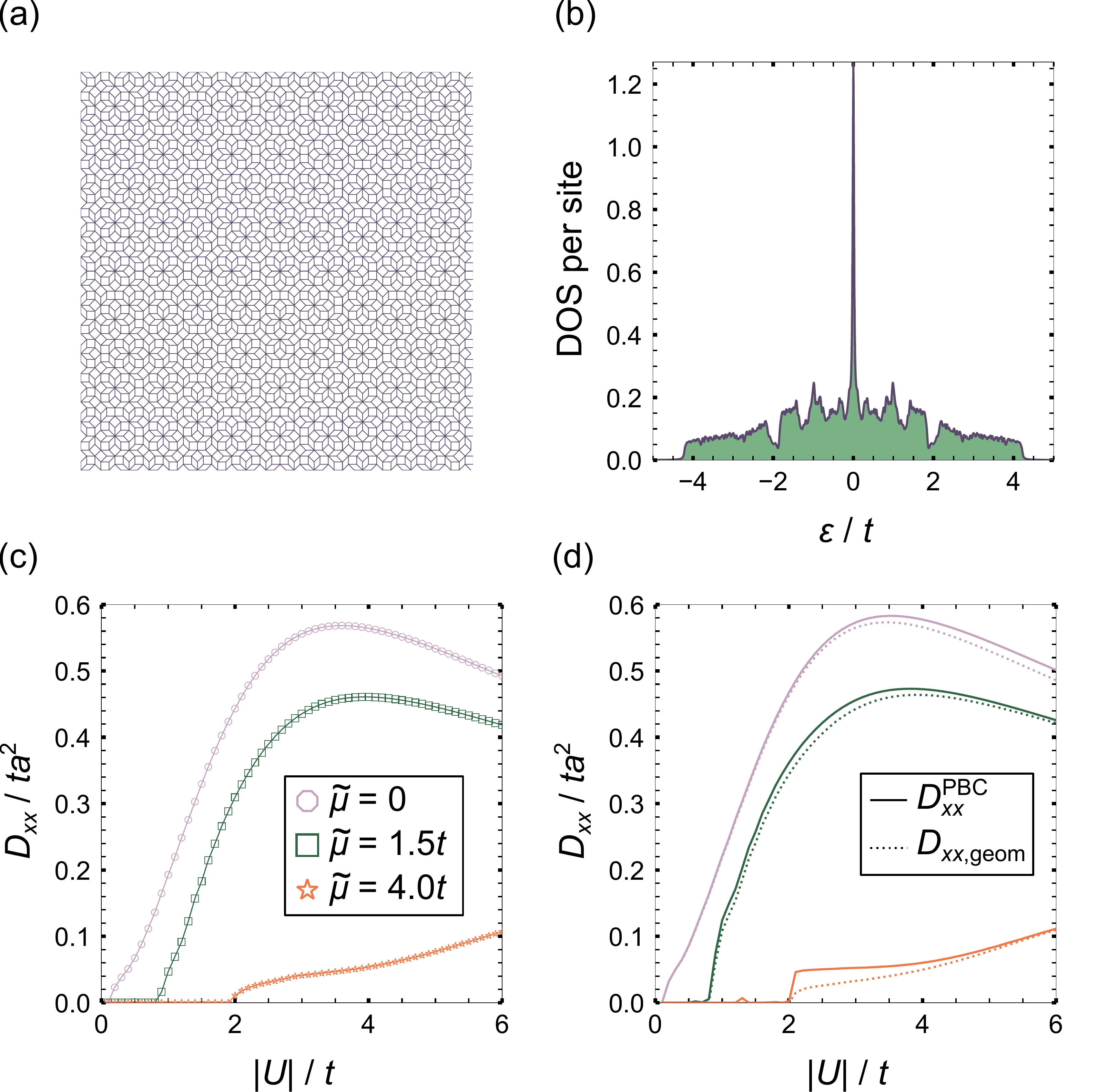}
    \caption{(a) The Ammann-Beenker quasicrystal with 2827 sites. Every vertex is the atomic site and each edge represents nearest-neighbor hopping link. We set the link length $a$ to unity for these calculations.
    (b) The DOS per site of the tight-binding Hamiltonian without the interaction.
    (c) and (d) Superfluid weight obtained with (c) open boundary and (d) periodic boundary conditions at zero temperature as a function of $|U| / t$ for each $\tilde{\mu}$.}
    \label{fig:AB}
\end{figure}

\textit{Ammann-Beenker quasicrystals.}
We investigate superfluid weight in quasiperiodic systems with critical states.
First, we study the Ammann-Beenker quasicrystal as a two-dimensional quasicrystal [Fig.~\ref{fig:AB}(a)].
The Hamiltonian on the Ammann-Beenker quasicrystal is written as
\begin{equation}
    \hat{H}_{\mathrm{AB}} = -t \sum_{\braket<j, l>} \sum_{\sigma} \hat{c}_{j\sigma}^{\dagger} \hat{c}_{l\sigma} - \mu \sum_{j\sigma} \hat{c}_{j\sigma}^\dagger \hat{c}_{j\sigma},
    \label{eq:H_QC}
\end{equation}
where $t$ is the hopping along nearest-neighbor links,
and $\mu$ is the chemical potential.
Here, $\braket<j, l>$ in the sum indicates the nearest-neighbor links.
We generate Ammann-Beenker quasicrystals with an inflation-deflation method with the periodic boundary condition~\cite{Ghadimi2021} along the x-axis as shown in Fig.~\ref{fig:AB}(a).

The eigenstates of the model in Eq.~\eqref{eq:H_QC} include critical and confined states. 
We show the DOS in Fig.~\ref{fig:AB}(b).
It is known that the states with the lowest and highest energy eigenvalues is critical~\cite{Mace2017}, which is neither extended nor exponentially localized.
In contrast, confined states are strictly localized states in a finite region.
All confined states lie at zero energy in this model~\cite{Araujo2019,Koga2020,Oktel2021}, leading to a peak of the DOS, which is similar to that of flat band systems.
Some numerical studies on the Ammann-Beenker quasicrystal have suggested that a large fraction of eigenstates except for confined states are critical~\cite{Grimmbook2002}.
Therefore, the geometric contribution to the superfluid weight is expected to become larger than the conventional one.

Figure~\ref{fig:AB}(c) shows $D_{xx}$ as a function of the onsite interaction $|U| / t$ obtained using Eq.~\eqref{eq:D_munu in BdGbasis} under open boundary conditions.
We note that the off-diagonal components of the superfluid weight tensor are vanishingly small
because the system is nearly four-fold rotational symmetric. 
The calculations are performed for several effective chemical potentials $\tilde{\mu} = \mu - \bar{V}$ renormalized by the Hartree potential.
For $\tilde{\mu} = 1.5t$ and $4.0t$, as self-consistent solutions with a finite order parameter could not be obtained below a certain value of $|U|$, $D_{xx}$ becomes finite above $|U| = 0.9$ and $2.0$, respectively.
For $\tilde{\mu} = 0$, where a strong peak lies in the DOS in the normal phase, $D_{xx}$ shows a linear dependence on the interaction $U$ when $U$ is weak.
This behavior coincides with that of the geometric superfluid weight in flat band superconductors~\cite{Peotta2015, Tovmasyan2016, Aleksi2016, Julku2020, Huhtinen2022}, which demonstrates that the superfluid weight in quasicrystals with confined states has a geometric origin, similar to that in periodic systems with flat bands.
Moreover, even though $\tilde{\mu}$ is shifted away from zero such as $\tilde{\mu} = 1.5t$ and $4.0t$, where confined states do not contribute, $D_{xx}$ shows a strong dependence on the interaction $U$.
This result suggests that the origin of the geometric superfluid weight in quasicrystals is not solely due to confined states but also related to critical states.

We also calculate the $|U|$ dependence of $D_{xx}^{\mathrm{PBC}}$ and $D_{xx,\mathrm{geom}}$ by imposing periodic boundary conditions in the $x$-direction.
The results, shown in Fig.~\ref{fig:AB}(d),
demonstrate that the geometric contribution is dominant for any $\tilde{\mu}$.
Notably, the calculations show $D_{xx}\simeq D_{xx, \mathrm{geom}}$,
as clearly seen in the case of $\tilde{\mu} = 4.0t$. 
This can be understood from the fact that the wavefunctions acquire a more critical character under the open boundary condition.
These results suggest that, while the geometric contribution is significant even in approximants, it becomes more dominant in true quasicrystals.

\textit{Aubry-Andr\'{e}-Harper model.}
The result in the Ammann-Beenker quasicrystal implies that critical states play an important role in the dominance of the geometric superfluid weight.
To clarify the reason, we investigate the Aubry-Andr\'{e}-Harper (AAH) model.
The Hamiltonian is described as a one-dimensional tight-binding model with a quasiperiodic potential~\cite{Aubry1980,Harper1955}, given by
\begin{align}
    \hat{H}_{\mathrm{AAH}} =& -t' \sum_{j \sigma} (\hat{c}_{j+1 \sigma}^{\dagger} \hat{c}_{j\sigma} + \hat{c}_{j\sigma}^{\dagger} \hat{c}_{j+1\sigma})
    \nonumber \\
    &+ \lambda \sum_{j} \cos \ab(\frac{2\pi}{\tau}ja) \hat{c}_{j\sigma}^{\dagger} \hat{c}_{j\sigma},
    \label{eq:H_AAH}
\end{align}
where $t'$ is a nearest-neighbor hopping, $\lambda$ is the strength of the quasiperiodic potential, $a$ is the lattice constant, and $\tau$ is the golden ratio.
For $\lambda = 0$, the model becomes a one-dimensional single-band model.
The AAH model exhibits a localization-delocalization transition at $\lambda =  2t'$.
At this transition point, all eigenstates are critical, and they exhibit power-law decay~\cite{Purkayastha2018, Sakai2022, Wang2025}.
For $\lambda < 2t' ~(\lambda > 2t')$, all eigenstates are extended (exponentially localized).
By approximating $\tau$ as an rational number on the model, the quasiperiodic potential can be smooth at the boundary with the periodic boundary condition.
Thus, we approximate $\tau$ as $987 / 610$ in this paper.

To confirm the phase transition, we show the energy spectrum and the multifractal dimension $D_2$ as a function of $\lambda / t'$ in Fig.~\ref{fig:AAH}(a).
See the Supplemental Materials~\cite{SupplementalMaterial} for the definition of $D_2$.
In the thermodynamic limit, $D_2 = 1 (0)$ for extended (exponentially localized) states,
whereas critical states give intermediate values~\cite{Fraxanet2022}.
Figure~\ref{fig:AAH}(a) shows the localization-delocalization transition and the emergence of critical states at $\lambda = 2t'$.

\begin{figure}[t]
    \centering
    \includegraphics[width=\columnwidth]{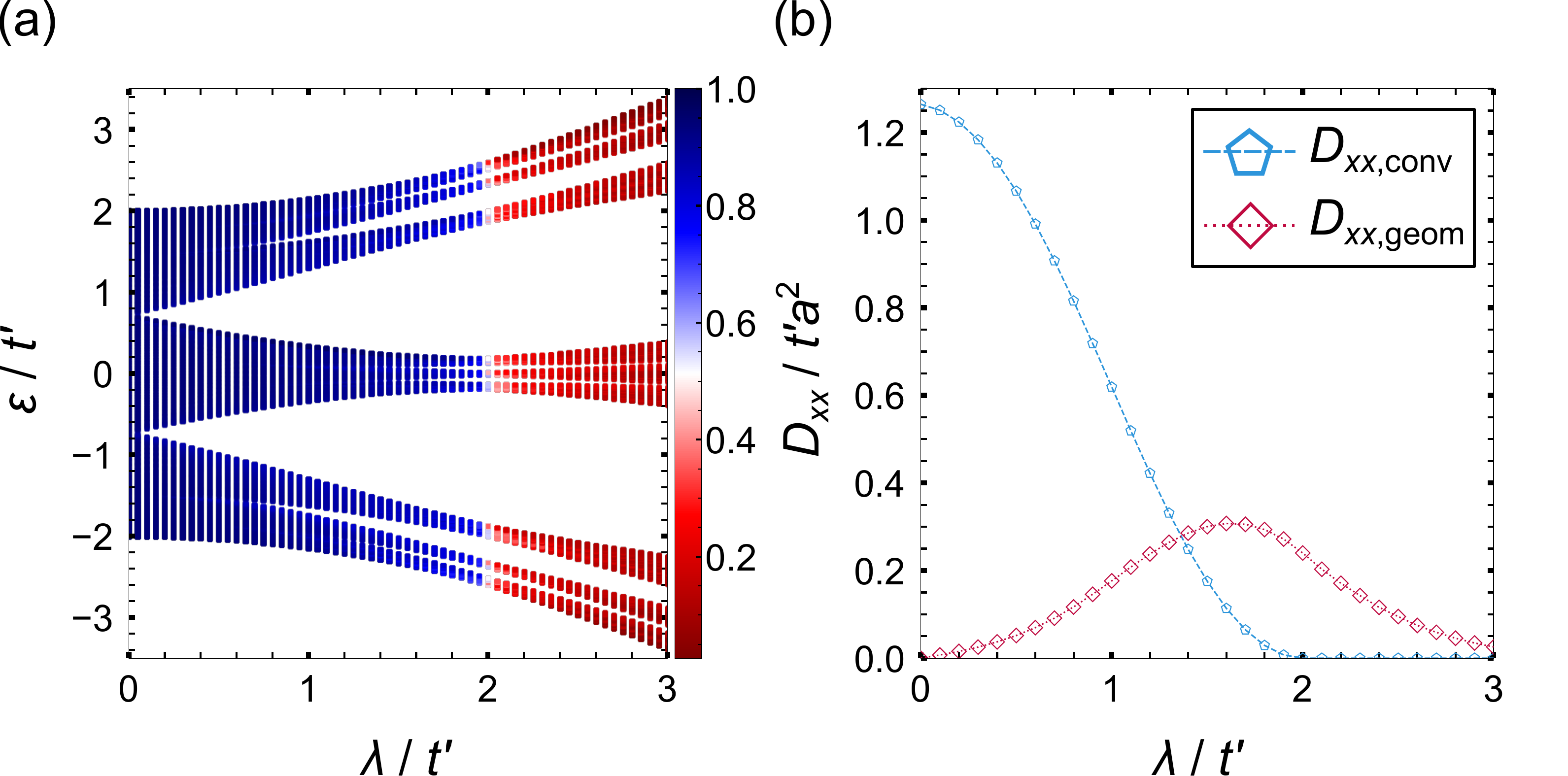}
    \caption{(a) Evolution of the energy spectrum of the AAH model. The color bar represents the multifractal dimension $D_2$. (b) Conventional superfluid weight $D_{xx, \mathrm{conv}}$ (blue) and geometric one $D_{xx, \mathrm{geom}}$ (red) for $U = -1.5t'$ at zero temperature.
    }
    \label{fig:AAH}
\end{figure}

We consider an $s$-wave superconductivity in the AAH model at half filling.
When the order parameter is allowed to be site-dependent, the localization-delocalization transition point depends on $U$~\cite{Zhang2022}.
For simplicity, we calculate the superfluid weight in the AAH model by assuming a uniform order parameter in order to fix the critical point to the analytical solution $\lambda = 2t'$.
The details of the self-consistent calculations are shown in the Supplemental Materials. 
Figure~\ref{fig:AAH}(b) shows conventional and geometric contributions of the superfluid weight as a function of $\lambda / t'$.
We first see the superfluid weight for $\lambda / t' < 2$.
While the conventional contribution decreases monotonically as the quasiperiodic potential $\lambda / t'$ increases, the geometric one becomes nonzero, and monotonically increases.
This increase of the geometric contribution is understandable from the fact that it is governed by interband matrix elements.
The interband contribution is absent at $\lambda = 0$ since the system is a single-band superconductor.
When the quasiperiodic potential is added, the single band is split due to broken translational symmetry, as shown in Fig.~\ref{fig:AAH}(a).
This allows various states with originally different wavevectors to be coupled, which can enhance the interband contribution~\cite{Alexander2022}.

Next, we focus on the superfluid weight at the transition point $\lambda = 2t'$.
As shown in Fig.~\ref{fig:AAH}(b), the conventional contribution vanishes at this point.
Namely, we find $D_{xx}^{\mathrm{PBC}} \simeq D_{xx, \mathrm{geom}}$ in the same way as in the Ammann-Beenker quasicrystal.
Our results indicate that the geometric contribution accounts for the superfluid weight when electronic states are critical.
Finally, for $\lambda \geq 2t'$, the contribution remains entirely geometric.
We note that the geometric contribution also decreases for $\lambda / t' > 2$,
because strong inhomogeneity due to quasiperiodicity suppresses the total superfluid weight.
The $|U|$ dependence obtained from the real-space and momentum-space approaches is shown in the Supplemental Material~\cite{SupplementalMaterial}.

\textit{Conclusion and discussion.}
In this paper, we have investigated the quantum-geometric superfluid weight in quasiperiodic superconductors.
We obtained the superfluid weight in the two-dimensional Ammann-Beenker quasicrystal and the Aubry-Andr\'{e}-Harper model, and revealed that the geometric contribution is enhanced by quasiperiodicity.
In the presence of critical states, the geometric contribution to the superfluid weight can be dominant.
Our result highlights the importance of geometric effects in understanding the origin of superconductivity in quasicrystals.

Additionally, a quasicrystalline potential has been experimentally realized in optical lattices~\cite{Viebahn2019, Sbroscia20}.
In quasicrystals with critical states, the superfluid weight shows linear dependence on the interaction strength due to the geometric effect.
Because the interaction is tunable in such systems, 
they would provide a platform for observing the geometric superfluid weight in quasicrystals.

\begin{acknowledgements}
    This work was supported by  JST SPRING (Grant
No. JPMJSP2151) and JSPS KAKENHI (Grant No. 23K13033, No. 24K00586 and 25H01248).
\end{acknowledgements}

\bibliography{QGQC}

\end{document}